\documentclass[letterpaper,english,reprint, aps]{revtex4-1}
\usepackage[T1]{fontenc}
\usepackage[latin9]{inputenc}
\usepackage{amsmath}
\usepackage{amssymb}
\usepackage{graphicx}

\makeatletter

\pdfpageheight\paperheight
\pdfpagewidth\paperwidth

\makeatother

\usepackage{babel}
\begin{document}
\preprint{APS/123-QED}
\title{Theoretical bound of the efficiency of learning with coarse-graining}
\author{Minghao Li}
\address{Department of Physics, Xiamen University, Xiamen, 361005, People's
Republic of China}
\author{Shihao Xia}
\address{Department of Physics, Xiamen University, Xiamen, 361005, People's
Republic of China}
\author{Youlin Wang}
\address{Department of Physics, Xiamen University, Xiamen, 361005, People's
Republic of China}
\author{Minglong Lv}
\address{Department of Physics, Xiamen University, Xiamen, 361005, People's
Republic of China}
\author{Shanhe Su}
\email{sushanhe@xmu.edu.cn}

\address{Department of Physics, Xiamen University, Xiamen, 361005, People's
Republic of China}
\date{\today}
\begin{abstract}
A thermodynamic formalism describing the efficiency of information
learning is proposed, which is applicable for stochastic thermodynamic
systems with multiple internal degree of freedom. The learning rate,
entropy production rate (EPR), and entropy flow from the system to
the environment under coarse-grained dynamics are derived. The Cauchy-Schwarz
inequality has been applied to demonstrate the lower bound on the
EPR of an internal state. The inequality of EPR is tighter than the
Clausius inequality, leading to the derivative of the upper bound
on the efficiency of learning. The results are verified in cellular
networks with information processes. 
\end{abstract}

\maketitle

\section{\textup{Introduction}}

Thermodynamic inequalities and uncertainties have a wide range of
applications in many fields {[}\citealp{lu2022geometric,pal2021thermodynamic,barato2015thermodynamic}{]}.
In this field, the role of information in thermodynamics are currently
an active research topic. The Landauer principle has stated that information
erasure requires heat dissipation \citep{landauer1961irreversibility},
which has been experimentally verified \citep{toyabe2010experimental,berut2012experimental}.
Markov approximation of the dynamics provides a versatile tool for
characterizing information in stochastic thermodynamic systems \citep{gingrich2017fundamental,lynn2022emergence,gingrich2016dissipation,lynn2022decomposing,lin2020three}.
The efficiency of information exchange was defined as the information
transfer over the total entropy production for a pair of Brownian
particles \citep{Allahverdyan_2009}. A rate of conditional Shannon
entropy reduction describing the learning of the internal process
about the external process has been proved to be bounded by the thermodynamic
entropy production {[}\citealp{barato2019unifying,barato2014efficiency,barato2013information}{]}
. 

Despite progresses in informational thermodynamics, one of the unresolved
questions is whether a fundemental bound of irreversible entropy production
exists in thermodynamic process involving information. Providing this
relationship would be beneficial for a better understanding of the
trade-off between the learning rate and the energy dissipation. 

In this work, the lower limit of the irreversible entropy production
rate of thermodynamic systems with multiple internal degrees of freedom
will be demonstrated and be applied to reveal the learning efficiency
for cellular networks. The contents are organized as follows: In Section
II, by using the Markovian master equation under coarse graining and
the Cauchy\textendash Schwartz inequality, an inequality associated
with the entropy flow from the system to the environment and the entropy
production rate is derived. A tighter upper bound for the learning
efficiency for one of the internal state is then obtained. In Section
III, through a single receptor model and an adaptive cellular network,
the learing efficiency for the biological information processing and
its upper bound could be verified. Finally, we draw our conclusions
in Sec. IV.

\section{\textup{Stochastic thermodynamics under coarse graining and learning
efficiency }}

We first briefly introduce the stochastic thermodynamics of information
processing system inspired by the Escherichia coli sensory network.
The evolution of the system is modeled by the Markovian master equation

\begin{equation}
\dot{p}(y,x)=\sum_{x^{\prime},y^{\prime}}\left[w_{x^{\prime}x}^{y^{\prime}y}p\left(y^{\prime},x^{\prime}\right)-w_{xx^{\prime}}^{yy^{\prime}}p\left(y,x\right)\right]
\end{equation}
where $p(y,x)$ is the probability of the system being in the discrete
state $(y,x)$, $x$ is an external state unaffected by the internal
state $y$, $w_{x^{\prime}x}^{y^{\prime}y}$ represents the transition
rate from state $(y^{\prime},x^{\prime})$ to state $(y,x)$. Note
that the overdot notation in this work is used to emphasize that the
quantity is a rate. Assuming that the external and internal states
never jump simultaneously, $w_{x^{\prime}x}^{y^{\prime}y}$ is simplied
as 

\begin{equation}
w_{x^{\prime}x}^{y^{\prime}y}=\begin{cases}
\begin{array}{c}
w_{x'x}^{y}\\
w_{y'y}^{x}\\
0
\end{array} & \begin{array}{c}
x\neq x',y=y'\\
x=x',y\neq y'\\
x\neq x',y\neq y',
\end{array}\end{cases}
\end{equation}
where $w_{x'x}^{y}$ is the transition rate of the external state
from $x'$ to $x$ given that the internal state is $y$, and similarly
for $w_{y'y}^{x}$. 

By considering an internal process comprising two variables $y=(y_{1},y_{2})$,
the transition rate $w_{y'y}^{x}=w_{(y_{1}',y_{2}')(y_{1},y_{2})}^{x}$.
The internal state $y_{1}$ may be indirectly connected to the external
state $x$ via the other internal state $y_{2}$. By using Eqs. (1)
and (2), the dynamics of the internal state $y_{1}$ given that the
external state takes $x$ is described by the master equation

\begin{equation}
\dot{p}\left(y_{1},x\right)=\sum_{y_{2}}\dot{p}\left(y_{1},y_{2},x\right)=\sum_{y_{1}'}J_{y_{1}'y_{1}}^{x},\label{eq:dynamic}
\end{equation}
where the marginal probability $p(y_{1},x)$ is obtained by summing
$p(y_{1},y_{2},x)$ over the variable $y_{2}$, $J_{y_{1}'y_{1}}^{x}=W_{y_{1}'y_{1}}^{x}p\left(y_{1}',x\right)-W_{y_{1}y_{1}'}^{x}p\left(y_{1},x\right)$,
and
\begin{equation}
W_{y_{1}y_{1}'}^{x}=\sum_{y_{2},y_{2}'}w_{(y_{1},y_{2})(y_{1}',y_{2}')}^{x}\frac{p\left(y_{1},y_{2},x\right)}{p\left(y_{1},x\right)}
\end{equation}
denotes the coarse-grained transition rate from state $(y_{1},x)$
to state $(y_{1}^{\prime},x).$

With the help of Eq. (\ref{eq:dynamic}), the time derivative of the
Shannon entropy $\dot{S}^{y_{1}}=-\sum_{y_{1}}\dot{p}\left(y_{1}\right)\mathrm{ln}p\left(y_{1}\right)$
of the internal state $y_{1}$ can be decomposed into three terms
{[}\citealp{su2022theoretical}{]}

\begin{equation}
\dot{S}^{y_{1}}=\dot{\sigma}^{y_{1}}-\dot{S}_{r}^{y_{1}}+\dot{l}^{y_{1}}\text{.}
\end{equation}
 $\dot{\sigma}^{y_{1}}$ and $\dot{S}_{r}^{y_{1}}$ are, respectively,
the thermodynamic entropy production rate and the entropy flow from
the system to the environment associated with the coarse-grained dynamics
of state $y_{1}$, which are defined as

\begin{equation}
\dot{\sigma}^{y_{1}}=\frac{1}{2}\sum_{y_{1},y_{1}',x}J_{y_{1}'y_{1}}^{x}\mathrm{ln}\frac{W_{y_{1}'y_{1}}^{x}p\left(y_{1}',x\right)}{W_{y_{1}y_{1}'}^{x}p\left(y_{1},x\right)},\label{eq:siga}
\end{equation}
and

\begin{equation}
\dot{S}_{r}^{y_{1}}=\frac{1}{2}\sum_{y_{1},y_{1}',x}J_{y_{1}'y_{1}}^{x}\mathrm{ln}\frac{W_{y_{1}'y_{1}}^{x}}{W_{y_{1}y_{1}'}^{x}}.\label{eq:sra}
\end{equation}
$\dot{\sigma}^{y_{1}}$ is always positive because of $W_{y_{1}'y_{1}}^{x}p(y_{1}',x)\geq0$
and $J_{y_{1}'y_{1}}^{x}\mathrm{ln}\frac{W_{y_{1}'y_{1}}^{x}p(y_{1}',x)}{W_{y_{1}y_{1}'}^{x}p(y_{1},x)}\geq0$
{[}15{]}.

\begin{equation}
\dot{l}^{y_{1}}=\frac{1}{2}\sum_{y_{1},y'_{1},x}J_{y'_{1}y_{1}}^{x}\mathrm{ln}\frac{p\left(x|y_{1}\right)}{p\left(x|y_{1}'\right)},\label{eq:la}
\end{equation}
quantifies the rate of the coarse-grained internal process $y_{1}$
learning about $x$ with $p(x|y_{1})=\frac{p(x,y_{1})}{p(y_{1})}=\frac{p(x,y_{1})}{\sum_{x}p(x,y_{1})}$
being the conditional probability.

Similar to Eq. (6), the thermodynamic entropy production rate, related
to the total internal state $y=\left(y_{1},y_{2}\right)$, is defined
as 
\begin{equation}
\dot{\sigma}^{y}=\frac{1}{2}\sum_{y,y',x}J_{y'y}^{x}\mathrm{ln}\frac{p\left(y',x\right)w_{y'y}^{x}}{p\left(y,x\right)w_{yy'}^{x}}
\end{equation}
with $J_{y'y}^{x}=w_{y'y}^{x}p\left(y',x\right)-w_{yy'}^{x}p\left(y,x\right)$.
The log-sum inequality $\sum_{i}u_{i}\mathrm{ln}\frac{u_{i}}{v_{i}}\geqslant\sum_{i}u_{i}\mathrm{ln}\frac{\sum_{i}u_{i}}{\sum_{i}v_{i}}$
$\left(\forall u_{i},v_{i}\geq0\right)$ {[}\citealp{zhen2021universal}{]}
implies that

\begin{equation}
\dot{\sigma}^{y}\geq\dot{\sigma}^{y_{1}}.
\end{equation}
This result can also be understood that the process of the change
of state $y_{1}$ is a part of the internal process.

In the cellular network, when $\dot{l}^{y_{1}}\geq0$, the cell creates
information by perceiving the change of the external state $x$. The
information learned will be consumed by the cell for driving the internal
process. Following Ref. {[}\citealp{goldt2017stochastic,barato2014efficiency,hartich2014stochastic}{]},
we define an learning efficiency $\eta^{y_{1}}$ for biological information
processing, which is given by
\begin{equation}
\eta^{y_{1}}=\frac{\dot{l}^{y_{1}}}{\dot{S}_{r}^{y_{1}}}.\label{eq:11}
\end{equation}
Under the condition of steady state, the change rate of the probability
$\dot{p}\left(y_{1},x\right)=0$, leading to the time derivative of
the Shannon entropy $\dot{S}^{y_{1}}=0$. Then, the EPR $\dot{\sigma}^{y_{1}}=\dot{S_{r}^{y_{1}}}-\dot{l}^{y_{1}}\geq0$
indicates that $\eta^{y_{1}}=\frac{\dot{l}^{y_{1}}}{\dot{S}_{r}^{y_{1}}}\leq1$.

By using the Cauchy\textendash Schwartz inequality $\sum_{i}a_{i}^{2}\sum_{i}b_{i}^{2}\geqslant\left(\sum_{i}a_{i}b_{i}\right)^{2}${[}\citealp{shiraishi2019fundamental}{]}
and the logarithmic inequality $\frac{(x-y)^{2}}{x+y}\leqslant\frac{x-y}{2}log\frac{x}{y}$
{[}\citealp{shiraishi2019fundamental}{]}, we derive the following
inequality associated with $\dot{S}_{r}^{y_{1}}$ and $\dot{\sigma}^{y_{1}}$ 

\begin{equation}
|\dot{S}_{r}^{y_{1}}|\leq\sqrt{\theta^{y_{1}}\dot{\sigma}^{y_{1}}}\label{eq:12}
\end{equation}
where $\theta^{y_{1}}=\frac{1}{2}\sum_{y_{1},y'_{1},x}\left(\mathrm{ln}\frac{W_{y_{1}'y_{1}}^{x}}{W_{y_{1}y_{1}'}^{x}}\right)^{2}W_{y_{1}'y_{1}}^{x}p\left(y_{1}',x\right)$.
In the case of $\dot{S}_{r}^{y_{1}}\geq0$ and $\dot{l}^{y_{1}}\geq0$,
a tighter upper bound for learning efficiency is obtained, i.e., 

\begin{equation}
\eta^{y_{1}}=\frac{\dot{l}^{y_{1}}}{\dot{S}_{r}^{y_{1}}}\leqslant1-\frac{\dot{S}_{r}^{y_{1}}}{\theta^{y_{1}}}.\label{eq:13}
\end{equation}

This bound, which is the main result of this work, is applicable to
plenty of systems as long as the principle of detailed balance is
satisfied. In the following section, the above results will be applied
to analysis cellular networks. By using the coarse-graining process,
the learing efficiency for biological information processing and its
upper bound could be determined. The costs and thermodynamic efficiencies
for various cellular networks learning about an external random environment
will be well characterized.

\section{\textup{The learning efficiencies of cellular networks }}

\subsection{Single receptor model}

In the cell, E. coli receptors are placed at the membrane and have
the ligand-binding site. The kinase is connected to the receptor and
its activity depends on the binding of external ligands. The kinase
in the active form acts as an enzyme of the phosphorylation reaction
of the protein. Here, we consider a single receptor accounting for
the indirect regulation of the kinase activity by the binding events.
The equivalent eight-state network of the single receptor model is
illustrated in Fig. 1(a). Each state of the system is described by
$(a,b,c)$. The internal process due to the change of the internal
state $y=(a,b)$ corresponds to a four-state network, as shown in
Fig. 1(b). The receptor is occupied by an external ligand at $b=1$
or unoccupied at $b=0$. For a given external ligand concentration
$c$, the free energy difference between the occupied and unoccupied
states is given by 
\begin{equation}
F\left(a,1,c\right)-F\left(a,0,c\right)=\ln\left(K_{a}/c\right),\label{eq:KAC}
\end{equation}
where $F(a,b,c)$ represents the free energy of state $(a,b,c)$,
$K_{a}$ is the dissociation constant, $\ln K_{a}$ is associated
with a change in the free energy of the receptor, and $\ln c$ denotes
the chemical potential of taking a particle from the solution in a
binding event {[}\citealp{barato2014efficiency}{]}. The subscript
in $K_{a}$ reflects the interaction between the receptor and the
kinase. The activity increases the dissociation constant by considering
that $K_{1}>K_{0}$. The state of the kinase attached to the receptor
may be inactive $a=0$ or active $a=1$, resulting in a conformational
change in the receptor. The free energy difference between the active
and inactive states with fixed values of $b$ and $c$ is assumed
to be $\Delta E$ , i.e., 
\begin{equation}
F\left(1,b,c\right)-F\left(0,b,c\right)=\Delta E.\label{DELTAe}
\end{equation}

Combining Eqs. (\ref{eq:KAC}) with (\ref{DELTAe}), we define the
free energy of state $(a,b,c)$ as

\begin{equation}
F\left(a,b,c\right)=a\Delta E+b\ln\left(K_{a}/c\right).\label{eq:free}
\end{equation}
The transition rate $w_{\left(a,b\right)\left(a',b^{\prime}\right)}^{c}$
from state $\left(a,b,c\right)$ to state $\left(a^{\prime},b^{\prime},c\right)$
given in Fig. 1(b) satisfies the detailed balance condition 
\begin{equation}
\ln\left[w_{\left(a,b\right)\left(a^{\prime},b^{\prime}\right)}^{c}/w_{\left(a^{\prime},b^{\prime}\right)\left(a,b\right)}^{c}\right]=F\left(a,b,c\right)-F\left(a^{\prime},b^{\prime},c\right).
\end{equation}
In Fig. 1(b), $\gamma_{a}$ and $\gamma_{b}$ are, respectively, the
time-scales of the conformational change and the binding event. The
timescale of the binding event is assumed to be much smaller than
that of the conformational changes, i.e., $\gamma_{b}\gg\gamma_{a}$.
In addition to the four-state subsystem in Fig. 1(b), the full model
includes the transition of the external ligand concentrations between
concentrations $c_{1}$ and $c_{2}$ at rate $\gamma_{c}$ {[}green
dashed arrows in Fig. 1(a) {]}. 

\begin{figure}
\includegraphics[viewport=80bp 50bp 500bp 490bp,clip,scale=0.35]{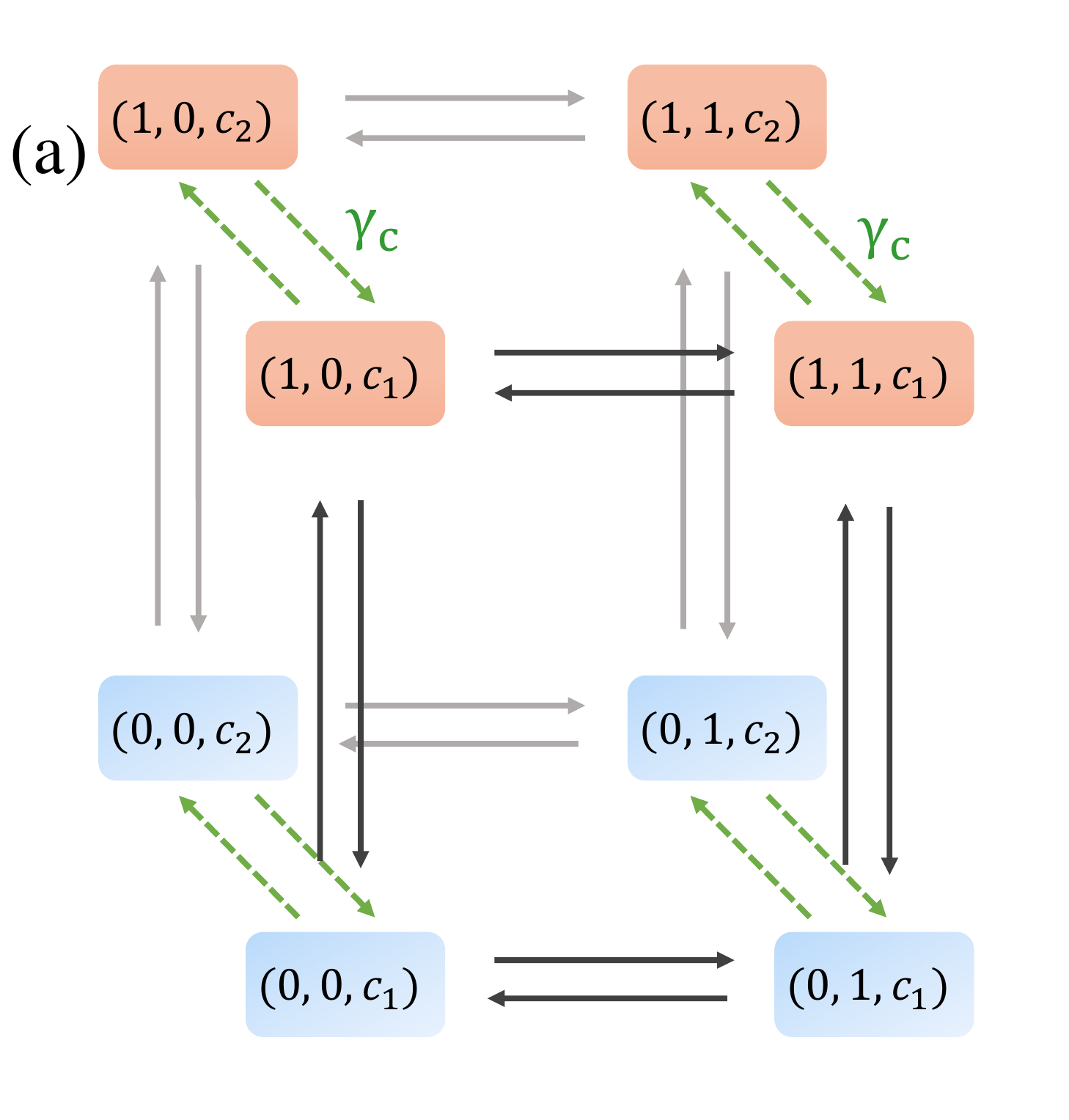}

\includegraphics[viewport=280bp 100bp 700bp 440bp,clip,scale=0.35]{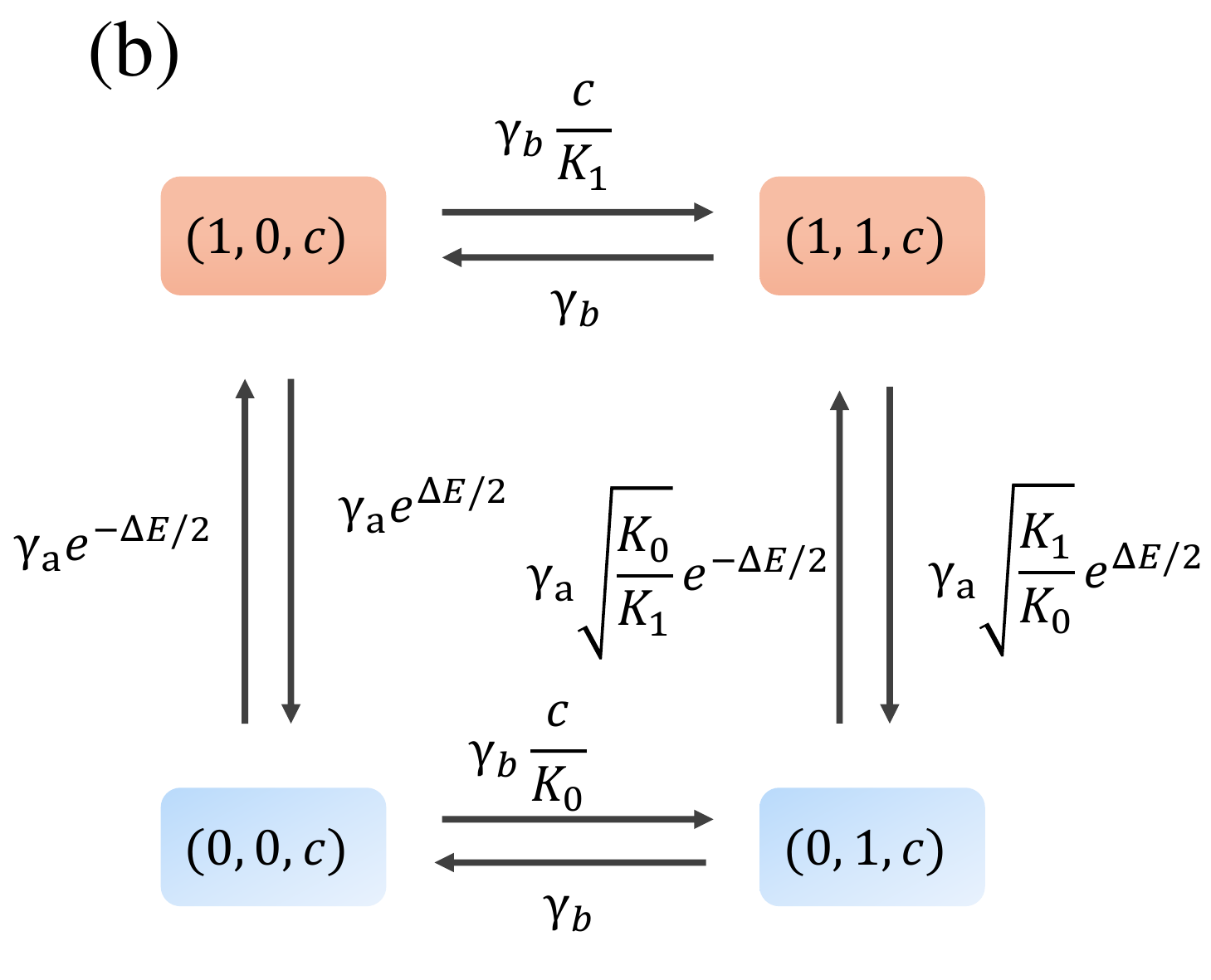}

\includegraphics[viewport=280bp 100bp 700bp 440bp,clip,scale=0.35]{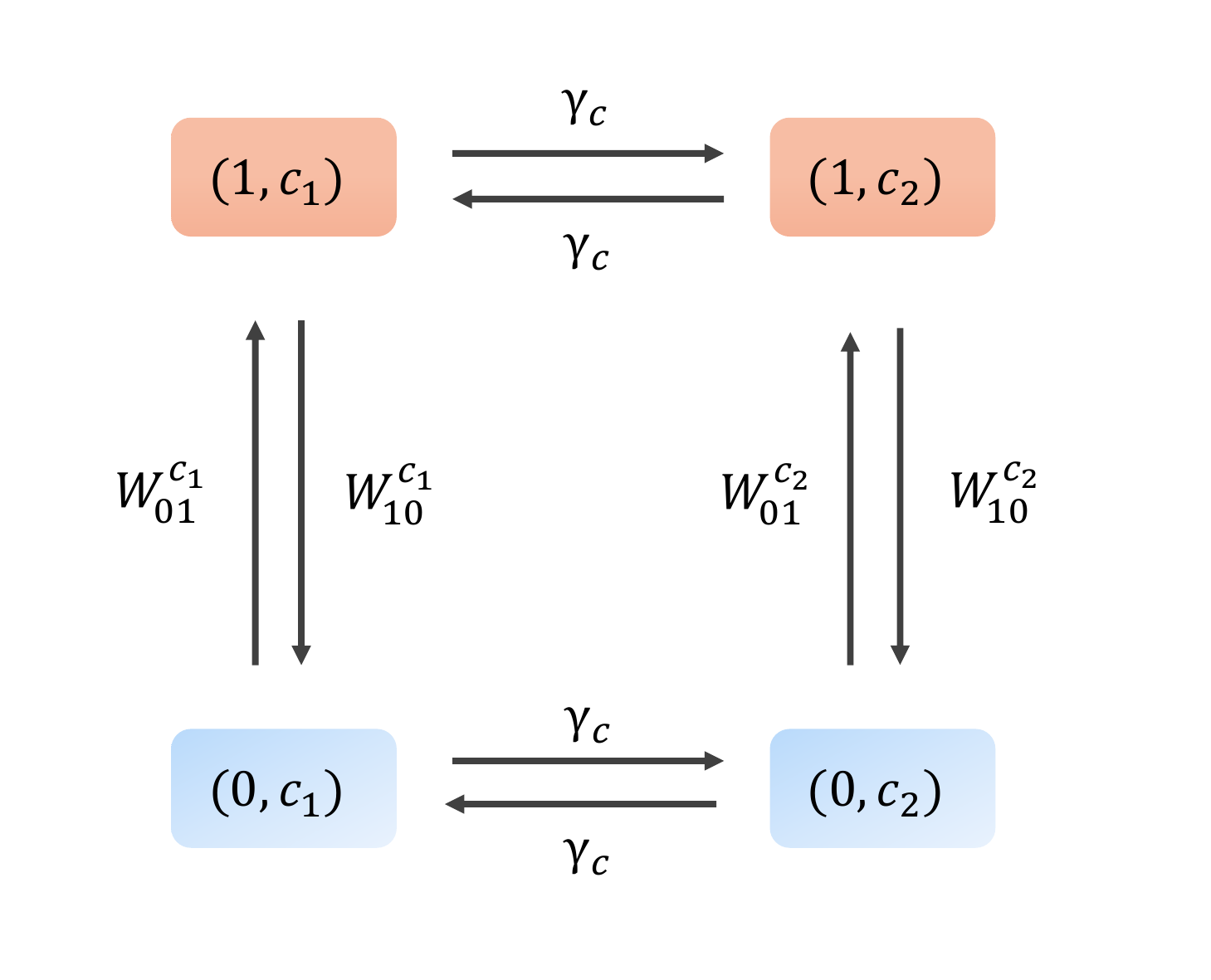}

\caption{(a) The schematic diagram of the single receptor model. (b) For a
given external ligand concentration of $c$, the transition rates
corresponding to the internal processes (the top subfigure). The schematic
diagram of the four-state coarse-graining model by summing out the
variable $b$ (the bottom subfigure). }
\end{figure}

By summing out the variable $b$, a coarse-grained model with four
states is shown by the bottom subfigure in Fig. 1 (b). The coarse-grained
trasition rate from state $(a,c)$ to $(a^{\prime},c)$ under the
external ligand concentration $c$ is given by 

\begin{equation}
W_{aa^{\prime}}^{c}=\sum_{b,b^{\prime}}w_{\left(a,b\right)\left(a^{\prime},b^{\prime}\right)}^{c}\frac{p\left(a,b,c\right)}{p\left(a,c\right)},\label{eq:Waa-1}
\end{equation}
where the probability $p\left(a,c\right)$ of coarsed-grained state
$\left(a,c\right)$ is calculated by the summation $\sum_{b}p\left(a,b,c\right)$.
In the limit $\gamma_{a}/\gamma_{b}\rightarrow0$, Eq. (\ref{eq:Waa-1})
is simplified as
\begin{equation}
W_{aa^{\prime}}^{c}=\sum_{b,b^{\prime}}w_{\left(a,b\right)\left(a^{\prime},b^{\prime}\right)}^{c}\frac{\left(c/K_{a}\right)^{b}}{1+c/K_{a}}\label{eq:Waa-1-1}
\end{equation}
with $\left(c/K_{a}\right)^{b}/\left(1+c/K_{a}\right)$ corresponding
to the stationary condition probability $p\left(b|a,c\right)=\frac{p\left(a,b,c\right)}{p\left(a,c\right)}$.
The free energy difference between state $\left(1,c\right)$ and state
$\left(0,c\right)$ due to the conformational change is then given
by

\begin{equation}
F\left(1,c\right)-F\left(0,c\right)=\ln\frac{W_{10}^{c}}{W_{01}^{c}}=\Delta E+\ln\left(\frac{1+\frac{c}{K_{0}}}{1+\frac{c}{K_{1}}}\right).\label{eq:F10C}
\end{equation}

According to Eqs. (\ref{eq:siga}) and (\ref{eq:sra}) and under the
stationarity condition, the entropy production rate associated with
the coarse-grained dynamics of state $a$ is given by 

\begin{align}
\dot{\sigma}^{a} & =\sum_{a,a^{\prime},c}W_{a^{\prime}a}^{c}p\left(a^{\prime},c\right)\mathrm{ln}\frac{W_{a^{\prime}a}^{c}p\left(a^{\prime},c\right)}{W_{aa^{\prime}}^{c}p\left(a,c\right)}\nonumber \\
 & =\dot{S}_{r}^{a}-\dot{l}^{a}\text{,}\label{eq:sig-1}
\end{align}
where the entropy flow from the cell to the environment related to
the change of $a$ reads

\begin{equation}
\dot{S}_{r}^{a}=\sum_{a,a^{\prime},c}W_{a^{\prime}a}^{c}p\left(a^{\prime},c\right)\mathrm{ln}\frac{W_{a^{\prime}a}^{c}}{W_{aa^{\prime}}^{c}},\label{eq:Sr-1}
\end{equation}
and the rate of the coarse-grained internal process $a$ learning
about the change of the external ligand concentration $c$ 
\begin{align}
\dot{l}^{a} & =-\sum_{a,a^{\prime},c}W_{a^{\prime}a}^{c}p\left(a^{\prime},c\right)\mathrm{ln}\frac{p\left(a^{\prime},c\right)}{p\left(a,c\right)}\nonumber \\
 & =\gamma_{c}\sum_{a}[p\left(a,c_{2}\right)-p\left(a,c_{1}\right)]\mathrm{ln}\frac{p\left(a,c_{2}\right)}{p\left(a,c_{1}\right)}.\label{eq:la-1}
\end{align}

By applying Eqs. (\ref{eq:11})-(\ref{eq:13}), the efficiency $\eta^{a}$
of the dynamics of the kinase in learning about the change of the
external ligand concentration 
\begin{equation}
\eta^{a}=\frac{\dot{l}^{a}}{\dot{S}_{r}^{a}}\leqslant1-\frac{\dot{S}_{r}^{a}}{\theta^{a}},\label{eq:eta-1}
\end{equation}
where the coefficient 

\begin{equation}
\theta^{a}=\frac{1}{2}\sum_{a,a',c}\left(\mathrm{ln}\frac{W_{a'a}^{c}}{W_{aa'}^{c}}\right)^{2}W_{a'a}^{c}p\left(a',c\right).\label{eq:the-1}
\end{equation}

\begin{figure}
\includegraphics[scale=0.17]{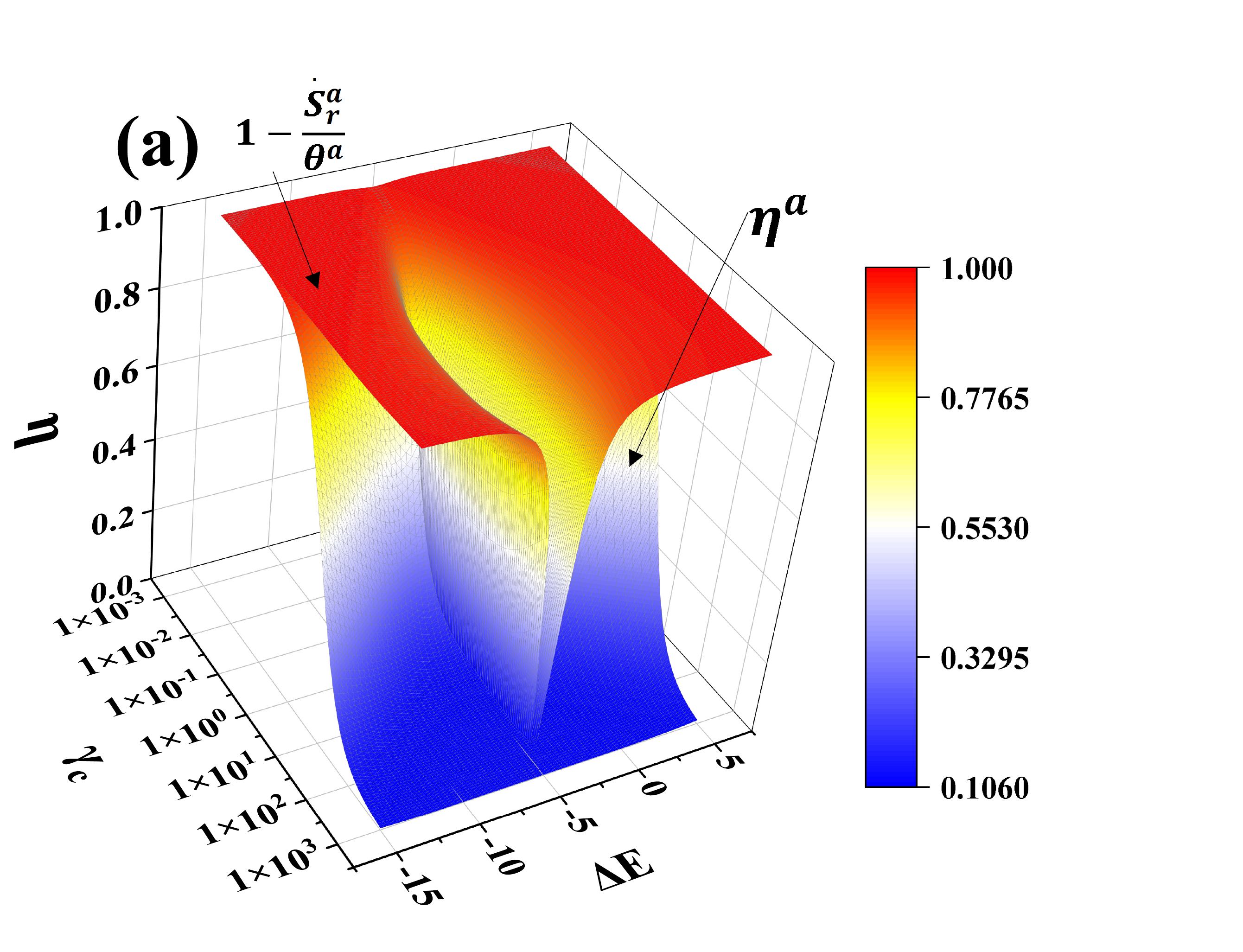}\includegraphics[scale=0.17]{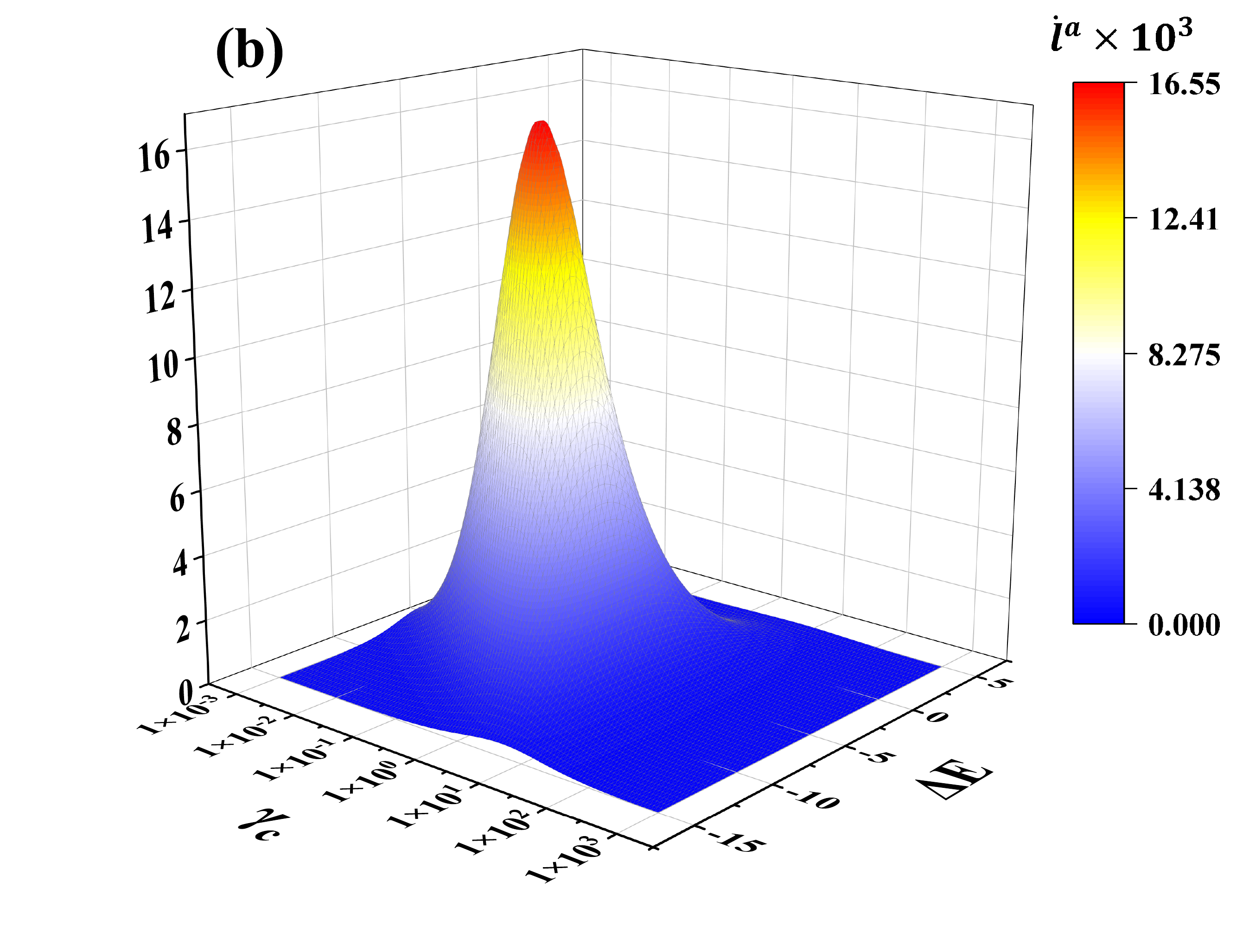}

\includegraphics[scale=0.17]{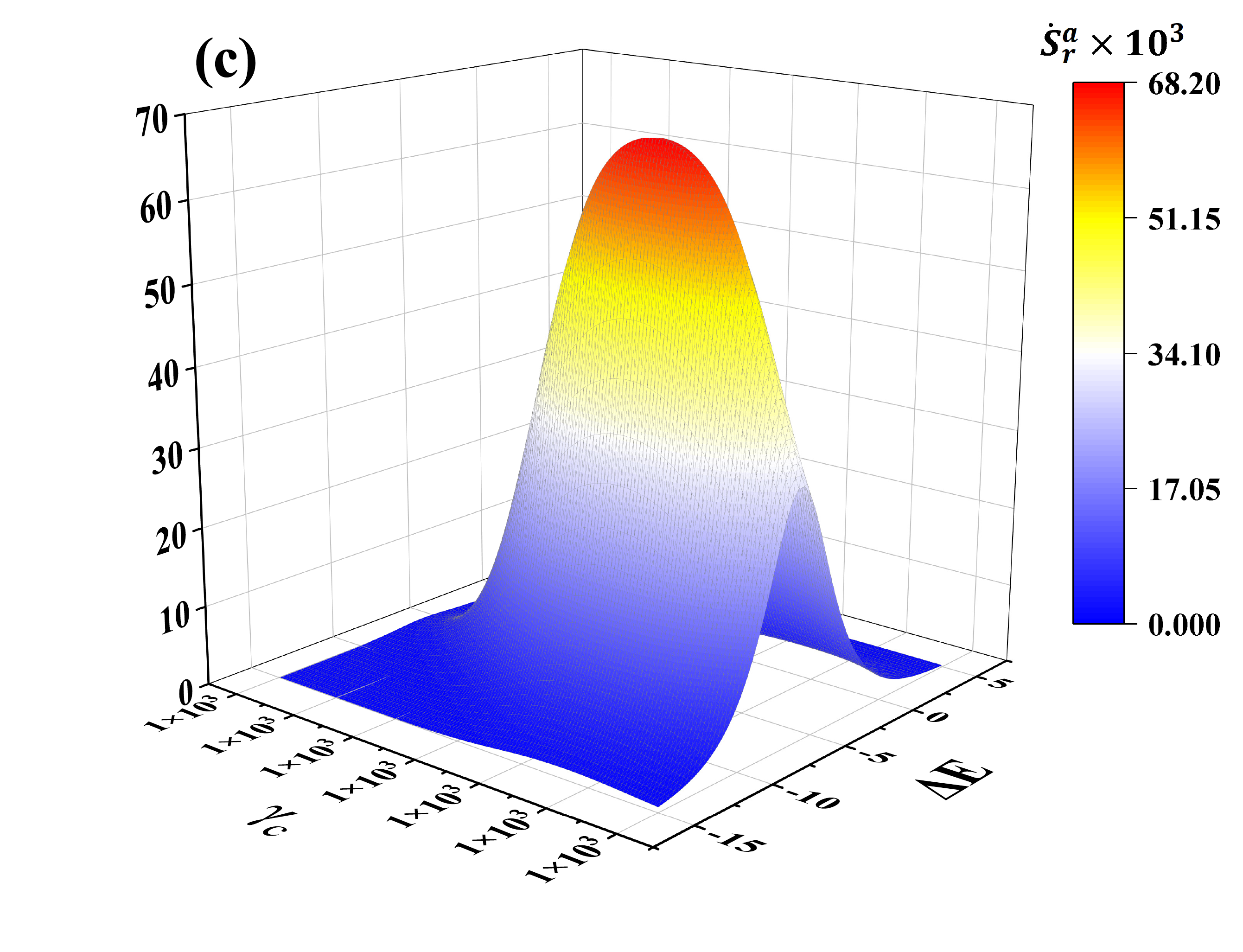}\includegraphics[scale=0.17]{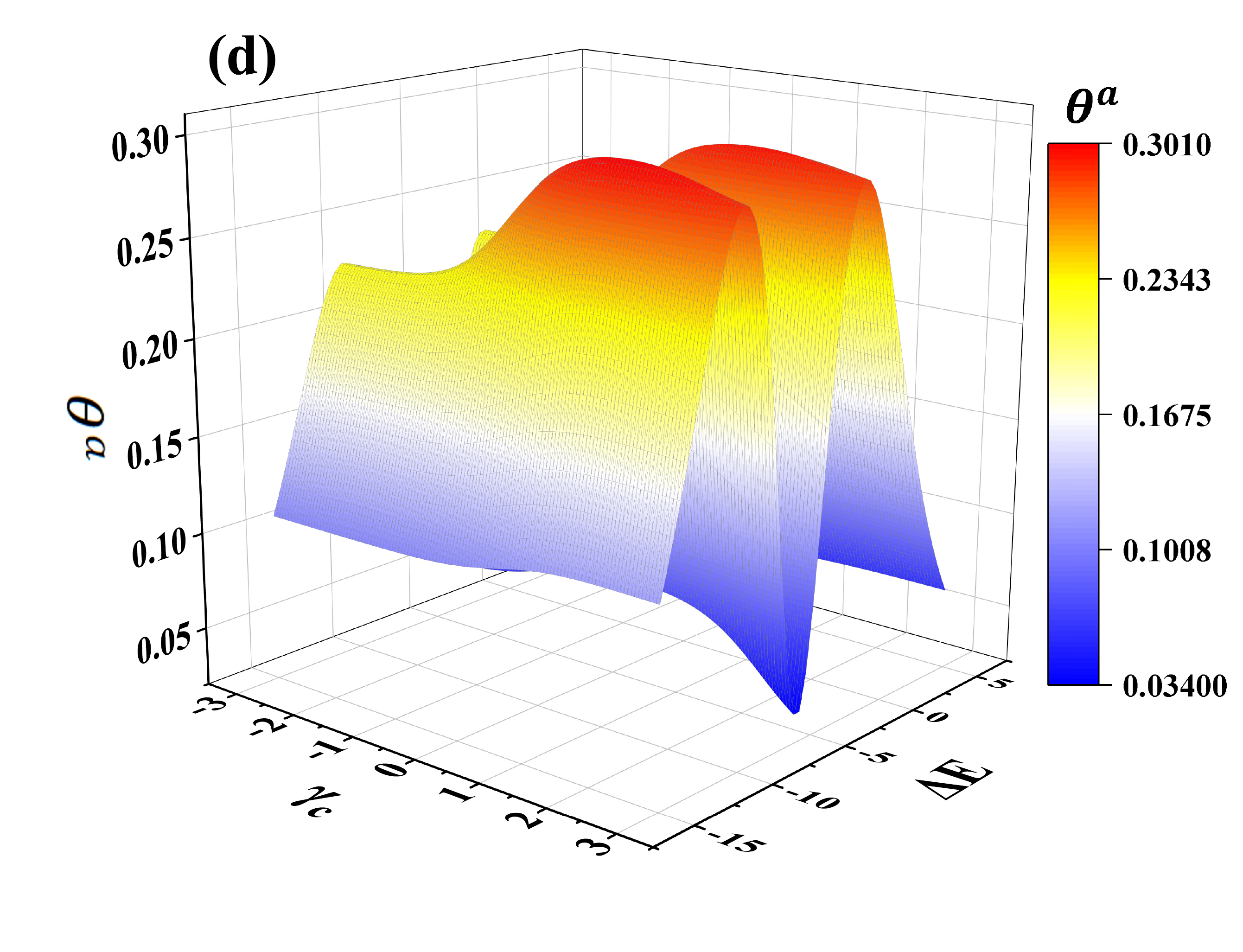}

\caption{(a) The learning efficiency $\eta^{a}$ and the upper limit $1-\frac{\dot{S}_{r}^{a}}{\theta^{a}}$
of the single receptor model varying with the transition rate $\gamma_{c}$
and the free energy difference $\Delta E$ between the active and
inactive states of CheA. (b) The learning rate $\dot{l}^{a}$ varying
with $\gamma_{c}$ and $\Delta E$. (c) The entropy flow $\dot{S}_{r}^{a}$
from the cell to the environment related to the change of $a$ varying
with $\gamma_{c}$ and $\Delta E$. (d) The coefficient $\theta^{a}$
varying with $\gamma_{c}$ and $\Delta E$. The other parameters $K_{0}=1/400$,
$K_{1}=400$, $\gamma_{a}=1,\gamma_{b}=1000,c_{1}=1/3$, and $c_{2}=3$. }
\end{figure}

By taking the limit $\gamma_{a}/\gamma_{b}\rightarrow0$ and using
Eq. (\ref{eq:Waa-1-1}), the entropy flow in (\ref{eq:Sr-1}) is simplied
as
\begin{equation}
\dot{S}_{r}^{a}=\gamma_{c}[p\left(0,c_{2}\right)-p\left(0,c_{1}\right)]\mathrm{ln}\left[\left(\frac{1+\frac{c_{1}}{K_{0}}}{1+\frac{c_{1}}{K_{1}}}\right)\left(\frac{1+\frac{c_{2}}{K_{1}}}{1+\frac{c_{2}}{K_{0}}}\right)\right],
\end{equation}
and the coefficient in (\ref{eq:the-1}) becomes
\begin{align}
\theta^{a} & =\frac{1}{2}\sum_{c}\gamma_{a}\left[\Delta E+\ln\left(\frac{1+\frac{c}{K_{0}}}{1+\frac{c}{K_{1}}}\right)\right]^{2}\left(1+\frac{c}{\sqrt{K_{0}K_{1}}}\right)\nonumber \\
 & \times\left(\frac{e^{\Delta E/2}p\left(1,c\right)}{1+c/K_{1}}+\frac{e^{-\Delta E/2}p\left(0,c\right)}{1+c/K_{0}}\right).\label{eq:the-1-yue}
\end{align}

Using Eqs. (\ref{eq:Sr-1})-(\ref{eq:the-1-yue}), the learning efficiency
$\eta^{a}$ varying with $\gamma_{c}$ and $\Delta E$ is plotted
in Fig. 2 (a). It is shown that the relation $\eta^{a}\leqslant1-\frac{\dot{S}_{r}^{a}}{\theta^{a}}$
always holds, which demonstrates the tightness of the bound in Eq.
(\ref{eq:eta-1}). When $\Delta E$ is fixed, $\eta^{a}$ decreases
monotonically with an increase in $\gamma_{c}$. If $\gamma_{c}\gg\gamma_{a}$,
the external concentration jumps too quickly that the intracellular
kinase activity couldn't accurately track the change in the external
environment, leading to the learning rate $\dot{l}^{a}\rightarrow0$
{[}Fig. 2 (b){]}. Consequently, $\eta^{a}$ is very small and deviates
from the upper bound $1-\frac{\dot{S}_{r}^{a}}{\theta^{a}}$ . When
$\Delta E$ is larger than $-2.2$ or smaller than $-12.5$, the entropy
flow $\dot{S}_{r}^{a}$ {[}Fig. 2 (c){]} and the coefficient $\theta^{a}$
{[}Fig. 2 (d){]} are not very sensitive to the variation of $\gamma_{c}$,
such that $1-\frac{\dot{S}_{r}^{a}}{\theta^{a}}$ does not change
significantly with respective to $\gamma_{c}$. For a $\Delta E$
between $-12.5$ and $-2.2$, the increase of $\gamma_{c}$ results
in the dramatic changes in the coefficient $\theta^{a}$ {[}Fig. 2
(d){]} and $1-\frac{\dot{S}_{r}^{a}}{\theta^{a}}$ {[}Fig. 2 (a){]}.
If $\gamma_{c}$ is much smaller than $\gamma_{a}$, both the actual
efficiency $\eta^{a}$ and the upper bound $1-\frac{\dot{S}_{r}^{a}}{\theta^{a}}$
approach unity, as the learning rate $\dot{l}^{a}$ {[}Fig. 2 (b){]}
and the entropy flow $\dot{S}_{r}^{a}$ {[}Fig. 2 (c){]} simultaneously
tend to zero. 

For a given value of $\gamma_{c}$, the values of the learning efficiency
$\eta^{a}$ and the upper bound $1-\dot{S}_{r}^{a}/\theta^{a}$ first
decrease and then increase with the increase of $\Delta E$ {[}Fig.
2 (a){]}. Compared with $\eta^{a}$, $1-\frac{\dot{S}_{r}^{a}}{\theta^{a}}$
is more sensitive to the change of $\Delta E$, because $\theta^{a}$
as a function of $\Delta E$ displays a bimodal structure with large
fluctuation {[}Fig. 2 (d){]}. There exists a point of $\Delta E$
where $1-\frac{\dot{S}_{r}^{a}}{\theta^{a}}$ reaches its minimum
value close to $\eta^{a}$ due to the fact that $\theta^{a}$ approaches
the lower limit {[}Fig. 2 (b){]} and $\dot{S}_{r}^{a}$ is relatively
large {[}Fig. 2 (c){]}. By varying with $\Delta E$, Figs. 2(b) and
2(c) indicate that both $\dot{l}^{a}$ and $\dot{S}_{r}^{a}$ will
have their respective maximum values.

\subsection{Model with adaptation}

Adaptation, in a biological sense, refers to a characteristic of an
organism that makes it fit for its environment. We consider an adaptive
model (Fig. 3) that incorporates the methylation level $m$ {[}\citealp{barato2014efficiency}{]},
which is a self-regulation factor for the activity $a$ of the kinase.
In this model, the activity $a=1$ if the kinase is active and $a=0$
if it is inactive, as shown in Fig. 3. Without the existence of external
ligands, the average value of $a$ is assumed to be equal to $1/2$
. A change in the external ligand concentration $c$ quickly changes
$a$ at a time-scale $\gamma_{\mathrm{a}}^{-1}$, while the methylation
level $m$ provides the adaption effect by adjusting the average of
$a$ back to $1/2$ at a time-scale $\gamma_{\mathrm{m}}^{-1}\gg\gamma_{\mathrm{a}}^{-1}$.
More specifically, a decreasing in $c$ leads to a fast growth of
$a$. The change in $a$ leads to a slow decrease in the methylation
level $m$ which acts back on the activity by slowly reducing $a$
to $1/2$. In the internal process of the cell, the methylation level
$m$ and the kinase activity $a$ forms a negative feedback network
that maintains $a$ at a stable level.

\begin{figure}
\includegraphics[scale=0.25]{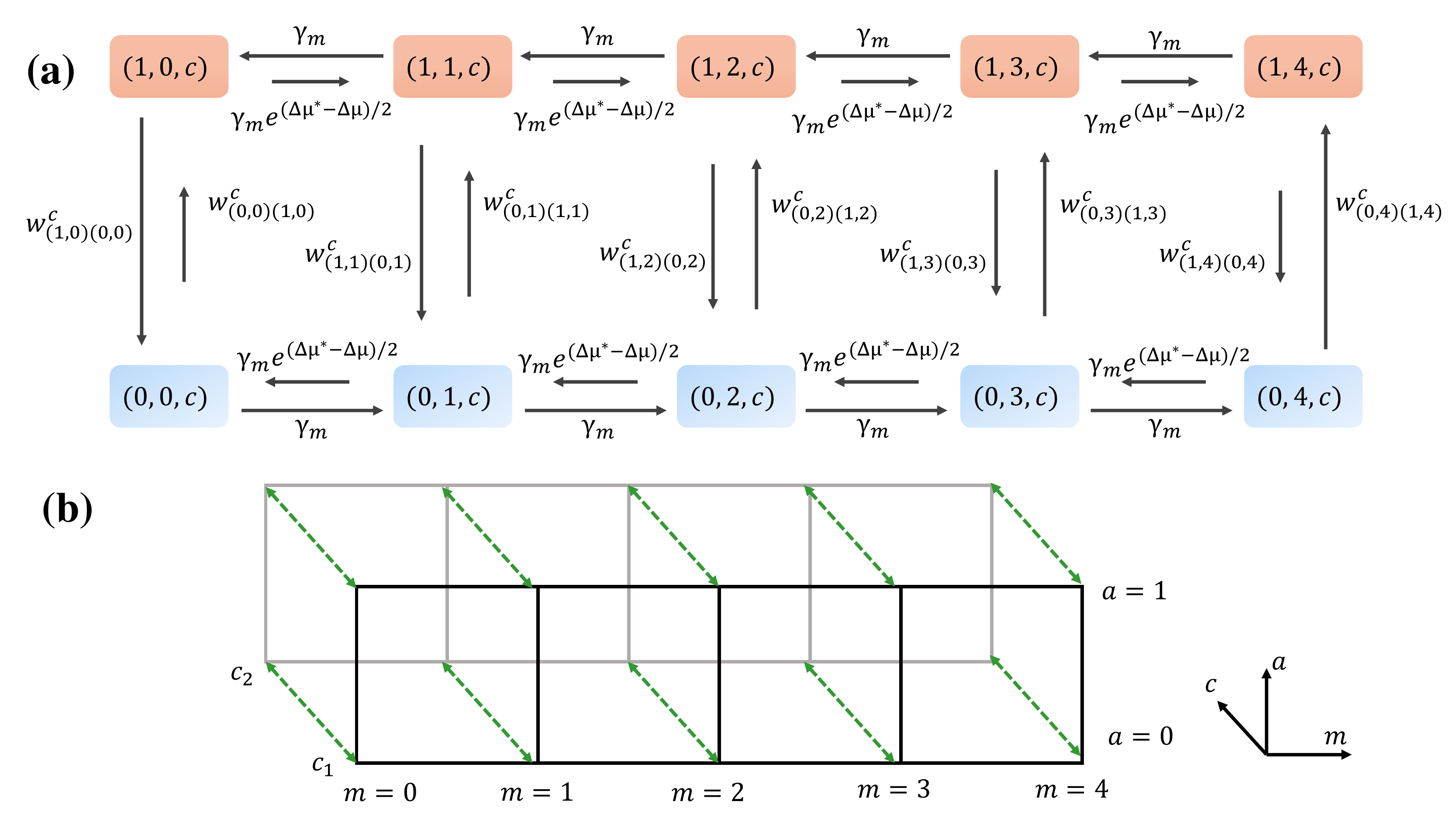}\caption{(a) The schematic diagram of the internal process in the adaptive
model. (b) The full model containing the jump of the external ligand
concentration.}
\end{figure}

By considering that the binding event is faster than the conformational
change and integrating out the variable $b$, the free energy difference
between the active state $(1,m,c)$ and the inactive state $(0,m,c)$
is obtained from Eq. (\ref{eq:F10C}), i.e.,
\begin{equation}
F\left(1,m,c\right)-F\left(0,m,c\right)=\Delta E\left(m\right)+\ln\left(\frac{1+\frac{c}{K_{0}}}{1+\frac{c}{K_{1}}}\right),\label{eq:fd}
\end{equation}
where the energy dependence on the methylation level
\begin{equation}
\Delta E\left(m\right)\equiv-\frac{m}{4}\ln\frac{K_{1}}{K_{0}}.\label{eq:Em}
\end{equation}
For simplicity, it is assumed that free energy of state $(a,m,c)$
\begin{equation}
F\left(a,m,c\right)=a\Delta E\left(m\right)+a\ln\left(\frac{1+\frac{c}{K_{0}}}{1+\frac{c}{K_{1}}}\right).
\end{equation}

The methylation level $m$ is controled by the chemical reactions.
If $a=0$, the receptor may be methylated due to the reaction 
\begin{equation}
[m]_{0}+\mathrm{SAM}\rightleftharpoons[m+1]_{0}+\mathrm{SAH},
\end{equation}
where SAM is a shortened form of the S-adenosyl methionine molecule
and SAH represents is an abbreviation of the S-adenosyl-L-homocysteine
molecule, and the subscript denotes the value of $a$. The free energy
difference of the reaction is given by $\mu_{\mathrm{SAM}}-\mu_{\mathrm{SAH}}$.
On the other hand, if $a=1$, the receptor may be demethylated with
the reaction 
\begin{equation}
[m+1]_{1}+\mathrm{H}_{2}\mathrm{O}\rightleftharpoons[m]_{1}+\mathrm{CH}_{3}\mathrm{OH},
\end{equation}
where the free energy difference
\begin{align}
\mu_{\mathrm{H}_{2}\mathrm{O}}-\mu_{\mathrm{CH}_{3}\mathrm{OH}}+F\left(1,m+1,c\right)-F\left(1,m,c\right)\nonumber \\
=\mu_{\mathrm{H}_{2}\mathrm{O}}-\mu_{\mathrm{CH}_{3}\mathrm{OH}}-\frac{1}{4}\ln\frac{K_{1}}{K_{0}}.
\end{align}

The chemical potential difference 
\begin{equation}
\Delta\mu=\mu_{\mathrm{SAM}}+\mu_{\mathrm{H}_{2}\mathrm{O}}-\mu_{\mathrm{SAH}}-\mu_{\mathrm{CH}_{3}\mathrm{OH}},
\end{equation}
provides the affinity for driving the internal process. 

The variable $(a,m)$ defines the internal process. The schematic
diagram of the internal process in the adaptive model and each transition
rate between difference states are shown in Fig. 3 (a). The full model
containing the jump of the external ligand concentration with rate
$\gamma_{c}$ between $c_{1}$ and $c_{2}$ is presented in Fig. 3
(b). 

In Fig. 3 (a), the trasition rate $w_{\left(1,m\right)\left(0,m\right)}^{c}$
from the active state $(1,m,c)$ and the inactive state $(0,m,c)$
is calculated by Eq. (\ref{eq:Waa-1-1}), while the only difference
is that the energy dependence on the methylation level in Eq. (\ref{eq:Em})
is taken into account. In addition, the trasition rates between state
$(1,m,c)$ and state $(0,m,c)$ is related to the free energy difference
in Eq. (\ref{eq:fd}) as
\begin{equation}
F\left(1,m,c\right)-F\left(0,m,c\right)=\ln\frac{w_{\left(1,m\right)\left(0,m\right)}^{c}}{w_{\left(0,m\right)\left(1,m\right)}^{c}}.
\end{equation}
The length of the arrow in Fig. 3 (a) indicates the relative magnitude
of $w_{\left(1,m\right)\left(0,m\right)}^{c}$ and $w_{\left(0,m\right)\left(1,m\right)}^{c}$.
The formulas of the trasition rates between state $(a,m,c)$ and state
$(a,m+1,c)$ are presented in Fig. 3 (a), ensuring that the adaptation
effects happens if $\Delta\mu$ overcomes the affinity in the internal
cycle $\Delta\mu^{*}=\frac{1}{4}\ln\frac{K_{1}}{K_{0}}$. In other
words, the increase of $a$ reduces the methylation level $m$ which
in turn tends to reduce the the value of $a$.

By using Eqs. (\ref{eq:sig-1})-(\ref{eq:la-1}), the coarse-graining
entropy production rate $\dot{\sigma}^{a}$ , entropy flow $\dot{S}_{r}^{a}$
from cell to environment, and learning rate $\dot{l}^{a}$ associated
with the change of the internal state $a$ in the adaptive model can
be computed. The only difference is that the coarse-grained trasition
rate from state $(a,c)$ to $(a^{\prime},c)$ under the external ligand
concentration $c$ is given by 

\begin{equation}
W_{aa^{\prime}}^{c}=\sum_{m,m^{\prime}}w_{\left(a,m\right)\left(a^{\prime},m^{\prime}\right)}^{c}\frac{p\left(a,m,c\right)}{p\left(a,c\right)},\label{eq:Waa-1-2}
\end{equation}
where the probability $p\left(a,c\right)$ of coarsed-grained state
$\left(a,c\right)$ is determined by the summation $\sum_{m}p\left(a,m,c\right)$.
In the same way, the efficiency $\eta^{a}$ of the dynamics of the
internal state $a$ in learning about the change of external ligand
concentration and its upper limit are computed by Eqs. (\ref{eq:eta-1})
and (\ref{eq:the-1}).

\begin{figure}
\includegraphics[scale=0.17]{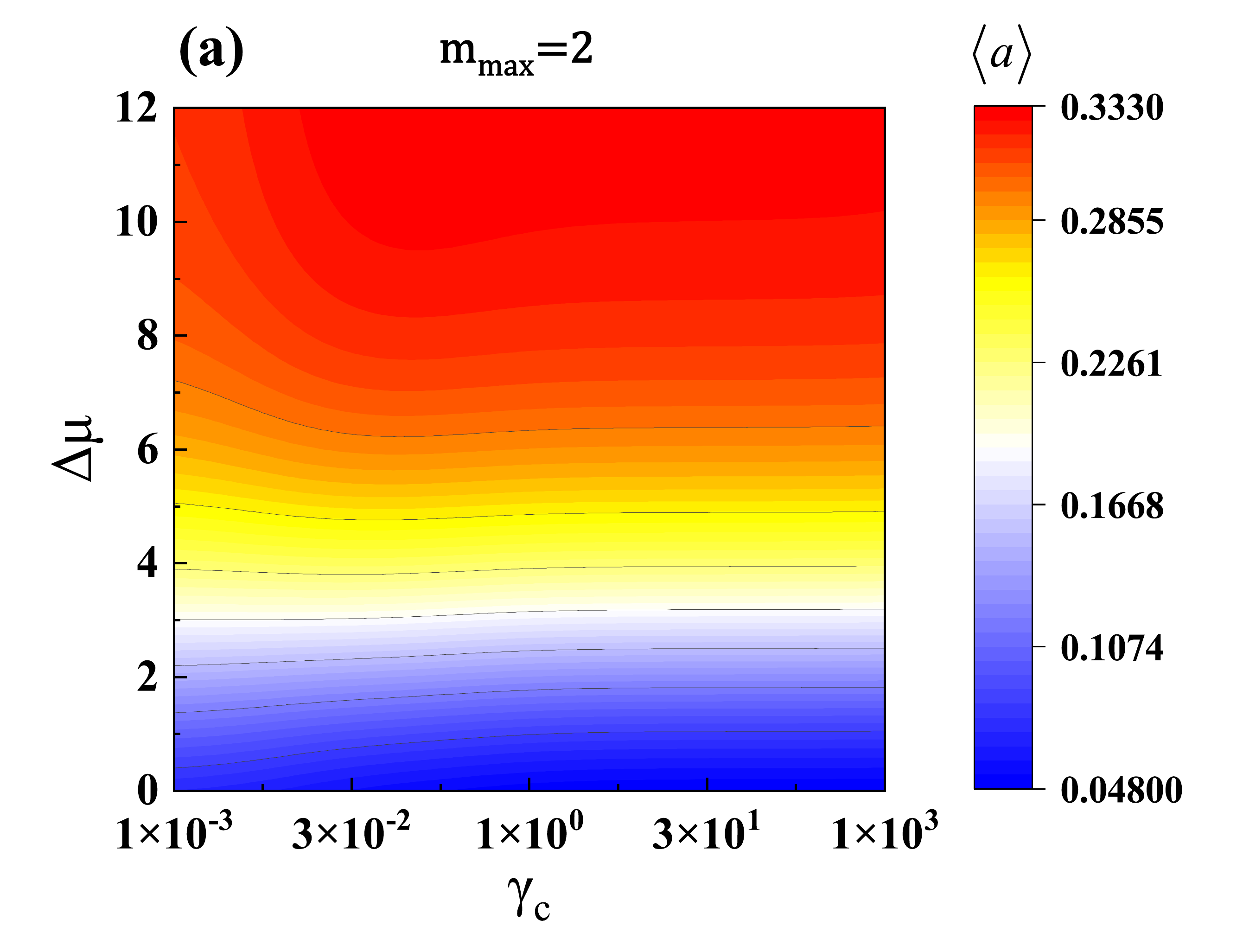}\includegraphics[scale=0.17]{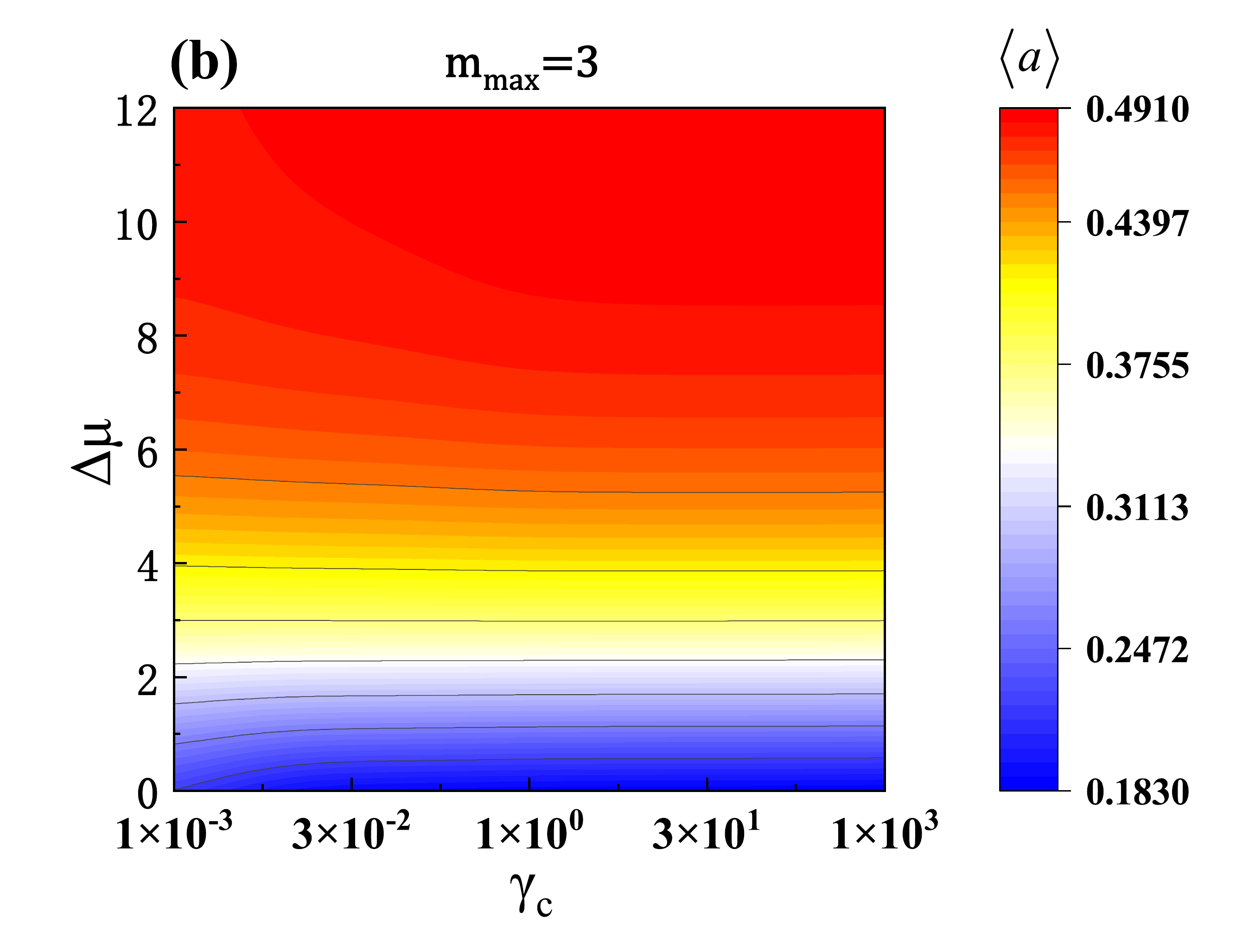}

\includegraphics[scale=0.17]{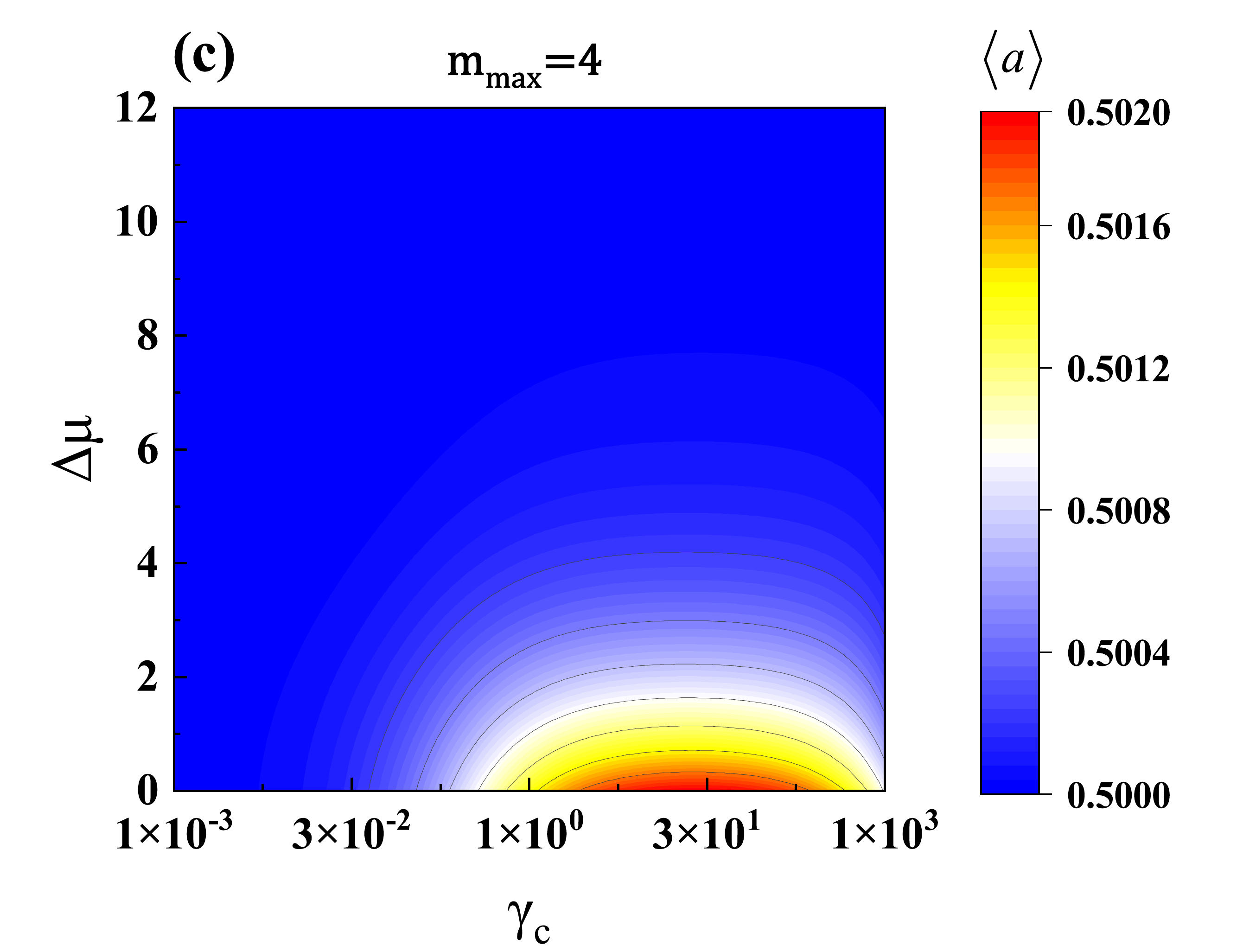}\includegraphics[scale=0.17]{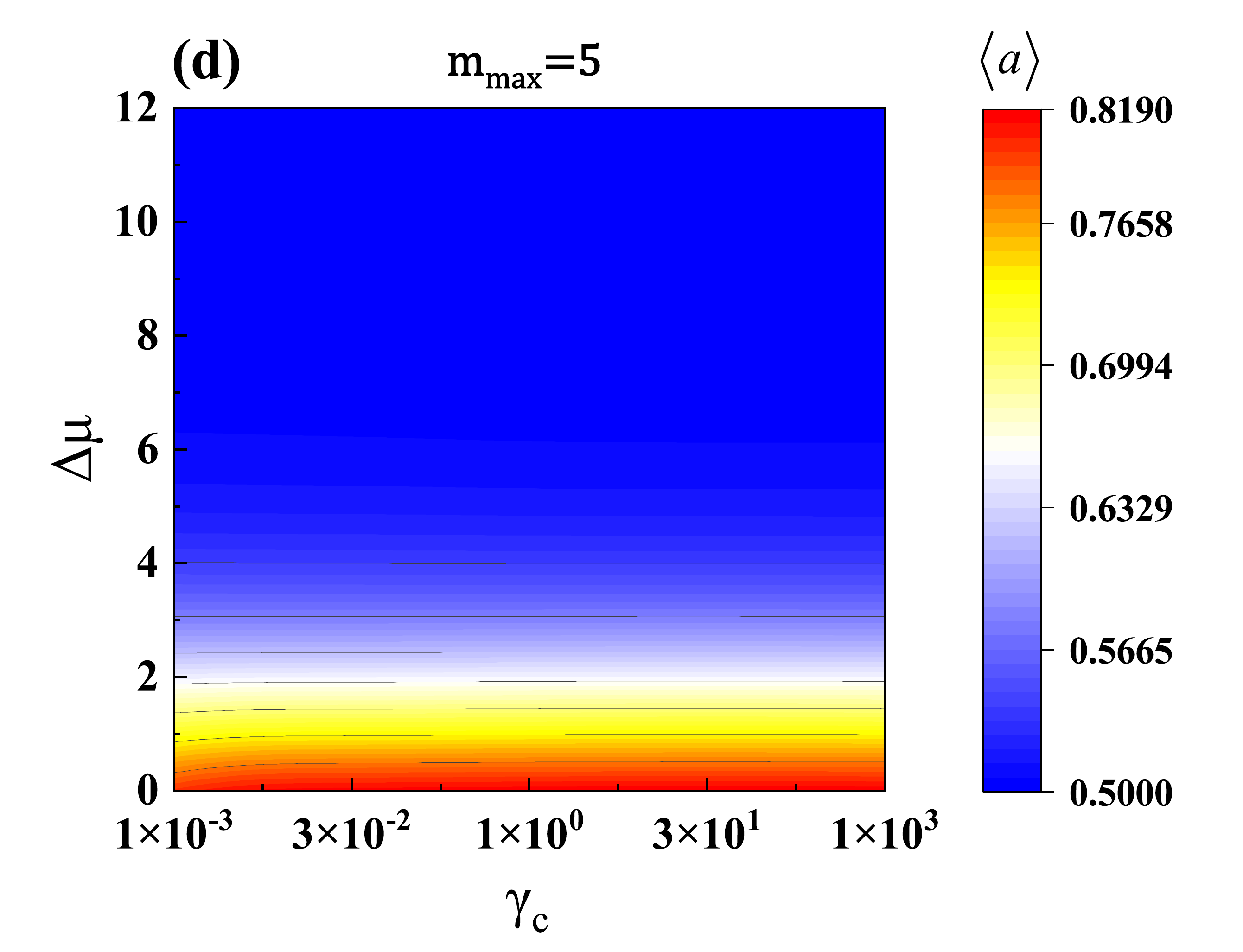}

\caption{For the maximum methylation level $m_{max}=2,3,4,5$, the average
$\left\langle a\right\rangle $ of state $a$ varying with the transition
rate $\gamma_{c}$ and the chemical potential difference $\Delta\mu$.
For the model with adaptation, the parameters $K_{0}=1/400$, $K_{1}=400$,
$\gamma_{a}=1,\gamma_{m}=10^{-2},c_{1}=1/3$, and $c_{2}=3$. }
\end{figure}

In the adaptive model, the intracellular kinase activity $a$ will
reponse to the change in the external ligand concentration $c$. This
process is indirectly regulated by altering the methylation level
$m$. For the maximum methylation level $m_{max}=2,3,4,5$, Fig. 4
plots the average $\left\langle a\right\rangle $ of state $a$ varying
with the transition rate $\gamma_{c}$ and the chemical potential
difference $\Delta\mu$. For a given level of methylation level $m$,
both $\gamma_{c}$ and $\Delta\mu$ affects the activity of the kinase.
When $m_{max}=4$, the average $\left\langle a\right\rangle $ of
state $a$ is more inclined to maintain around $1/2$ even with large
varivations of $\gamma_{c}$ and $\Delta\mu$. That is to say that
the feedback network in Fig (3) is capable of maintaining $a$ at
a stable level by optimizing the methylation level $m$.

By applying Eqs. (\ref{eq:Sr-1})-(\ref{eq:the-1}) and choosing $m_{max}=4$,
Fig. 5 plots the learning efficiency $\eta^{a}$ and the upper bound
$1-\frac{\dot{S}_{r}^{a}}{\theta^{a}}$ for the adaptive model varying
 with the transition rate $\gamma_{c}$ and the chemical potential
difference $\Delta\mu$. As illustrated in Fig. 5 (a), for a fixed
value of $\Delta\mu$, $\eta^{a}$ and its upper bound $1-\frac{\dot{S}_{r}^{a}}{\theta^{a}}$
decrease monotonically as $\gamma_{c}$ increases, while the learning
rate $\dot{l}^{a}$, entropy flow $\dot{S}_{r}^{a}$, and coefficient
$\theta^{a}$ appear to be nonmonotonic functions of $\gamma_{c}$.
When $\gamma_{c}$ is fixed, $\eta^{a}$ and $1-\frac{\dot{S}_{r}^{a}}{\theta^{a}}$
are not sensitive to the change of $\Delta\mu$, as $\dot{l}^{a}$,
$\dot{S}_{r}^{a}$, and $\theta^{a}$ monotonicaly increase with the
increase of $\Delta\mu$ at the same time. The increase of $\Delta\mu$
will accelerate the transition rates between different methylation
levels, which makes the response of $m$ to the change in the external
concentration $c$ more sensitive. This enables the internal state
$a$ learning more external information per unit time, but also makes
the network also generates more dissipation. It is important to observe
that $\eta^{a}$ is always smaller than $1-\frac{\dot{S}_{r}^{a}}{\theta^{a}}$,
which shows the evidence about the validity of Eq. (\ref{eq:eta-1}). 

\begin{figure}
\includegraphics[viewport=30bp 0bp 772bp 591bp,scale=0.17]{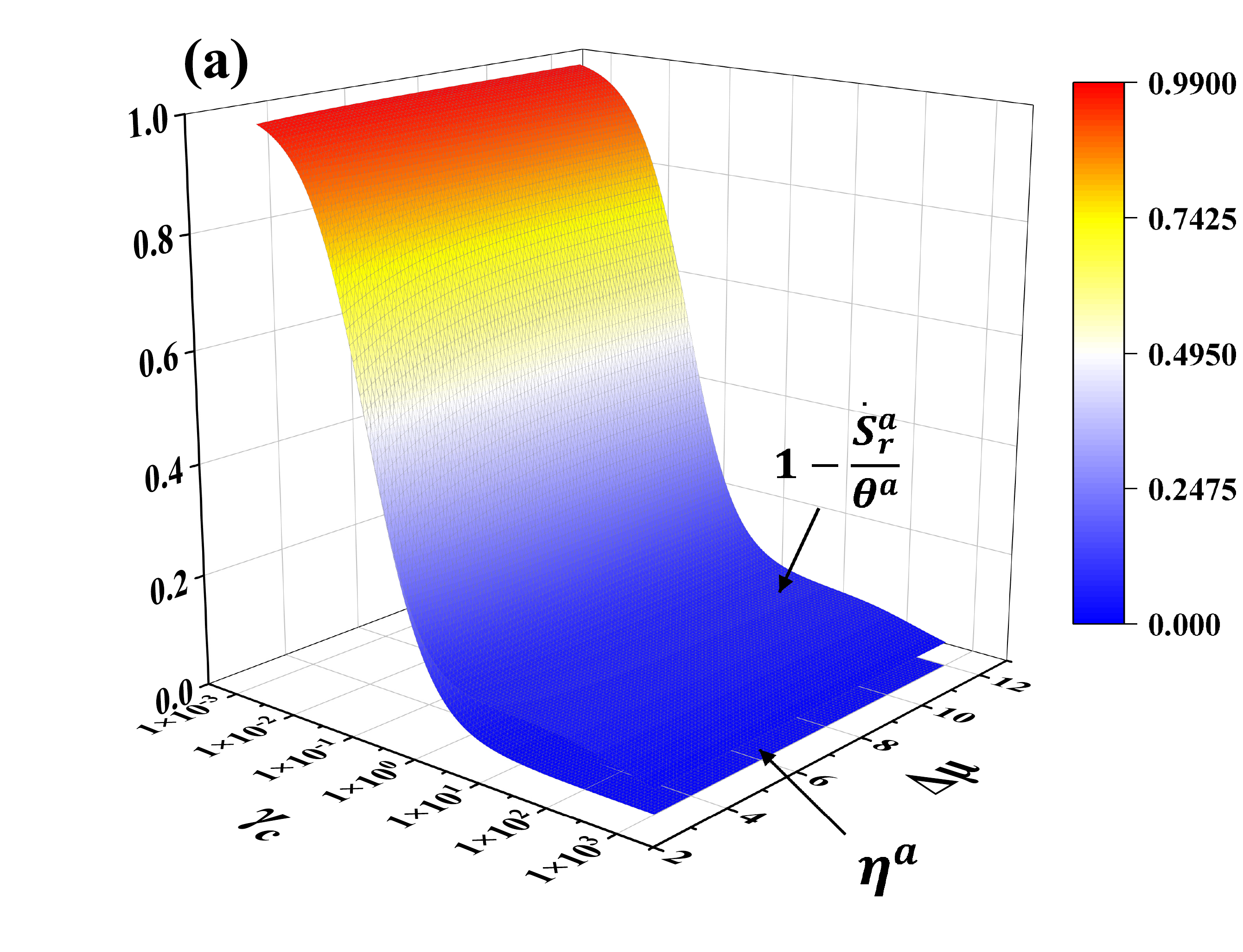}\includegraphics[scale=0.17]{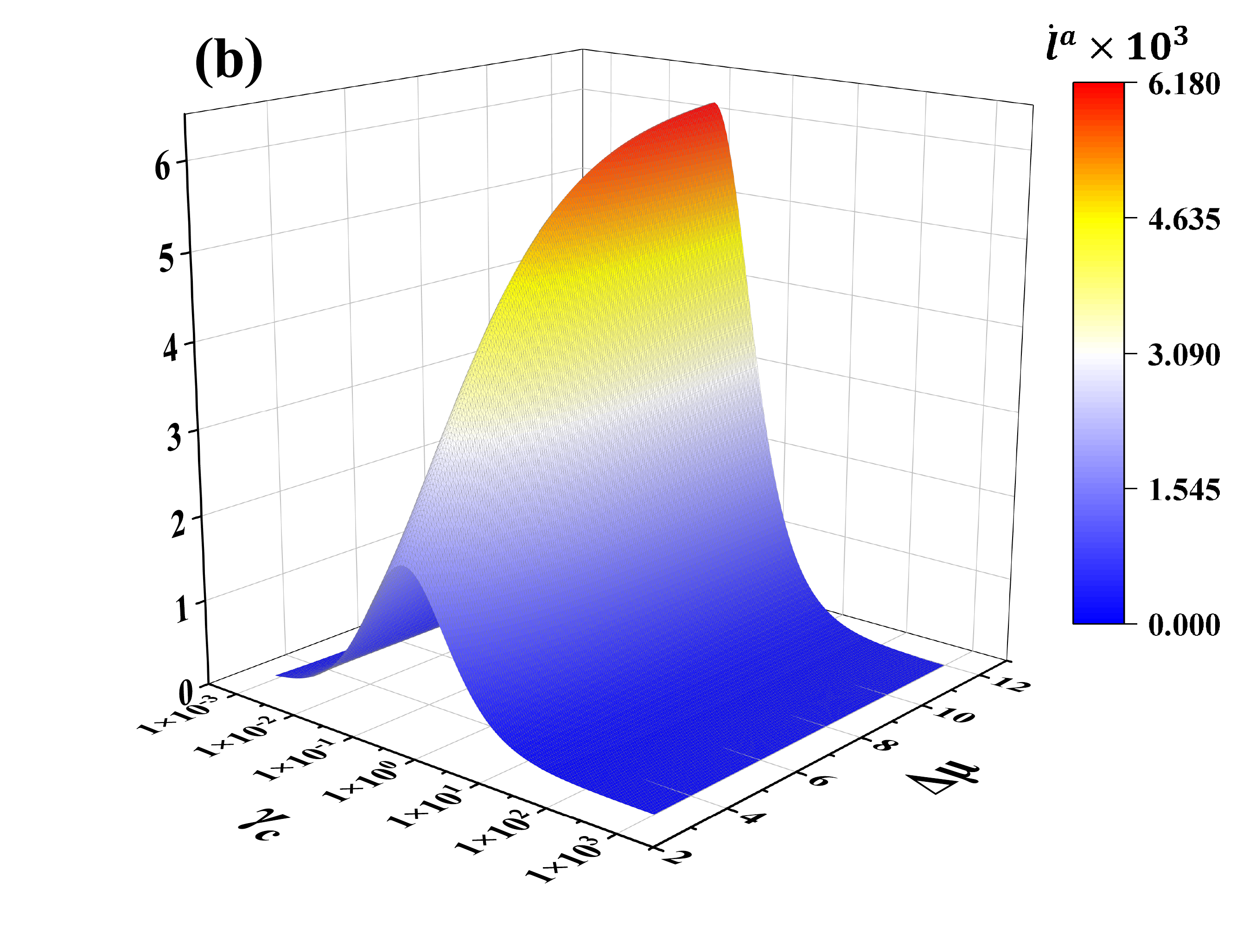}

\includegraphics[scale=0.17]{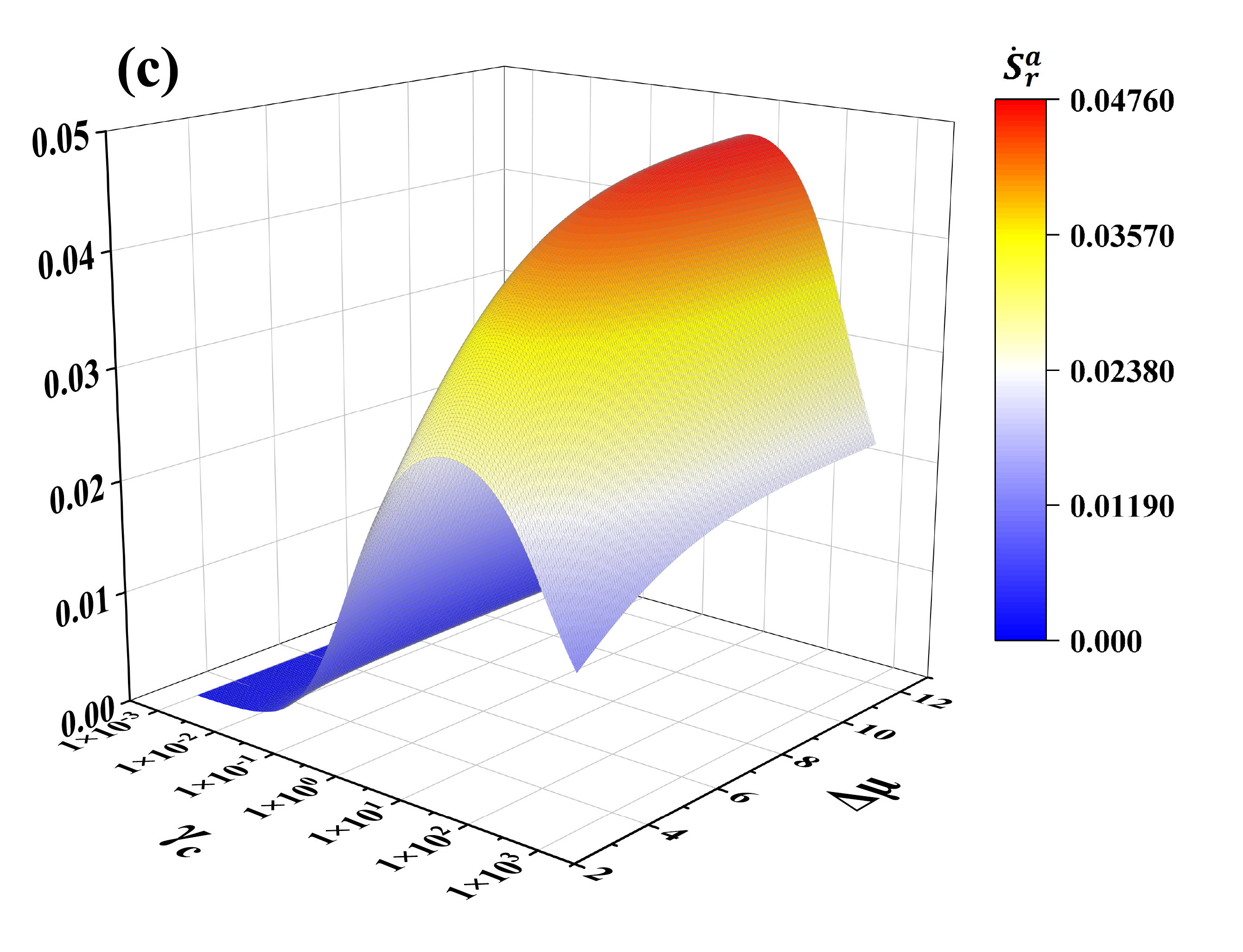}\includegraphics[scale=0.17]{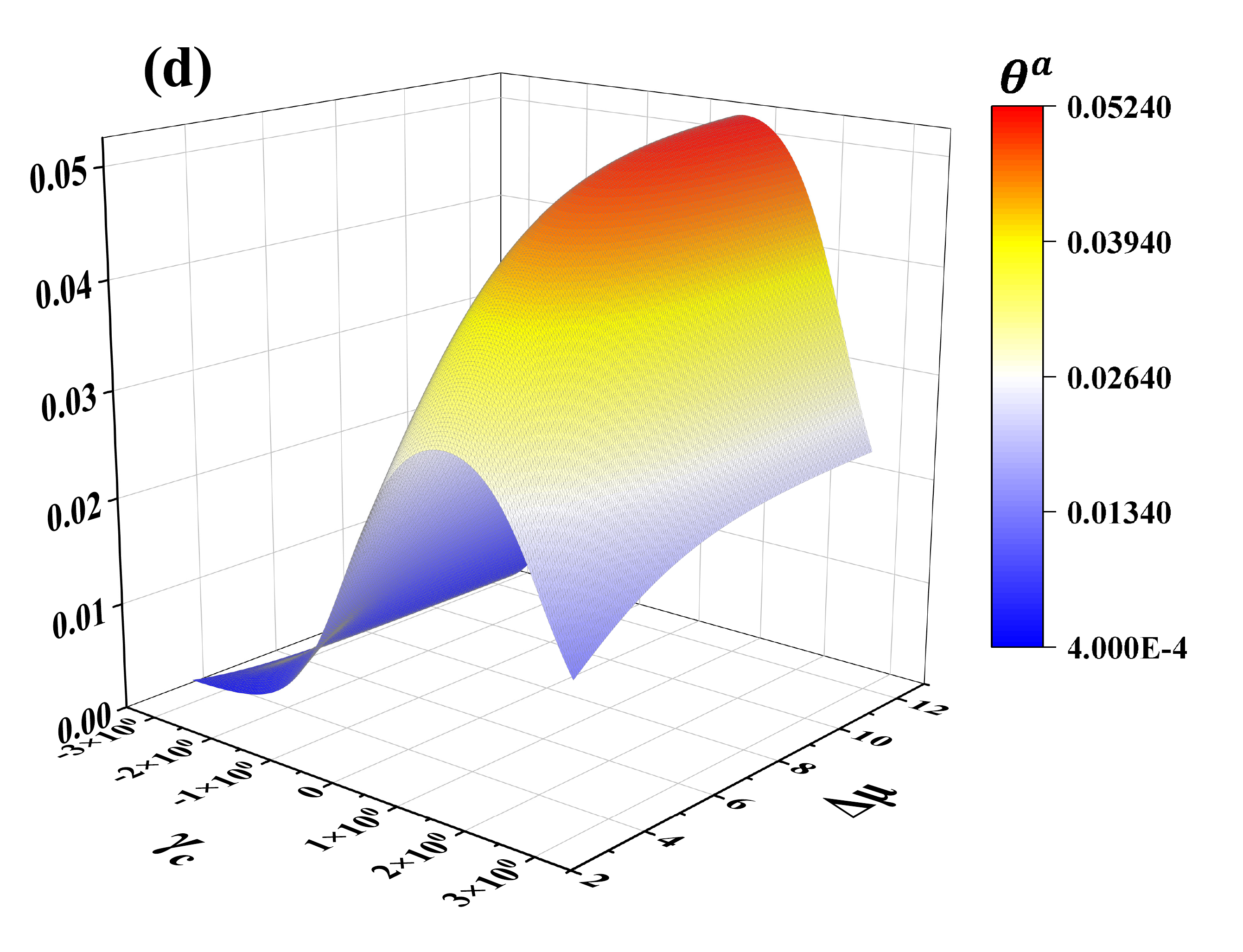}

\caption{(a) The learning efficiency $\eta^{a}$ and the upper limit $1-\frac{\dot{S}_{r}^{a}}{\theta^{a}}$
of the model with adaptation varying with the transition rate $\gamma_{c}$
and chemical potential difference $\Delta\mu$. (b) The learning rate
$\dot{l}^{a}$ varying with $\gamma_{c}$ and $\Delta\mu$. (c) The
entropy flow $\dot{S}_{r}^{a}$ from the cell to the environment related
to the change of $a$ varying with $\gamma_{c}$ and $\Delta\mu$.
(d) The coefficient $\theta^{a}$ varying with $\gamma_{c}$ and $\Delta\mu$.}
\end{figure}

\section{\textup{Conclusion}}

A coarse-graining approach has been developed to quantify the upper
bound of the efficiency of learning in multivariable systems, which
is tighter than that given by the conventional second law of thermodynamics.
By applying this approach to thermodynamic processes in cellular networks,
the general applicability of our method has been demonstrated. Our
findings have important implications for understanding the fundamental
principles governing the dynamics of complex systems.
\begin{acknowledgments}
This work has been supported by the National Natural Science Foundation
(Grants No. 12075197) and the Fundamental Research Fund for the Central
Universities (No. 20720210024). 
\end{acknowledgments}

\bibliographystyle{apsrev4-1}
\nocite{*}
\bibliography{ref1}

\end{document}